%% file: ms.tex
\documentclass[12pt,preprint]{aastex}

\newcommand{\m}{$^{m}~~$}

\newcommand{\be} {\begin{equation}}

\newcommand{\ee} {\end{equation}}
\newcommand{\etal}{{\em et al.} }

\def\rxj {1RXS\,J170849$-$400910}
\def\ee {1E\,2259$+$586}

\newcommand{\BSAX}{{\em Beppo}SAX} 
\newcommand{\RXTE}{{\em R}XTE}
\newcommand{\bc}{\begin{center}}
\newcommand{\ec}{\end{center}}
\def\ltsima{$\; \buildrel < \over \sim \;$}
\def\lsim{\lower.5ex\hbox{\ltsima}}
\def\loe{\lower.5ex\hbox{\ltsima}}
\def\gtsima{$\; \buildrel > \over \sim \;$}
\def\gsim{\lower.5ex\hbox{\gtsima}}
\def\goe{\lower.5ex\hbox{\gtsima}}

\shortauthors{REA ET AL.}
\shorttitle{CYCLOTRON FEATURE IN 1RXS J170849$-$400910}

\begin{document}
\title{EVIDENCE FOR A CYCLOTRON FEATURE IN THE SPECTRUM OF THE ANOMALOUS X-RAY 
PULSAR 1RXS J170849$-$400910}

\author{ Nanda Rea\altaffilmark{1,2}, Gian Luca Israel\altaffilmark{2,7}, Luigi
  Stella\altaffilmark{2,7}, Tim Oosterbroek\altaffilmark{3}, Sandro
  Mereghetti\altaffilmark{4}, Lorella Angelini\altaffilmark{5},
  Sergio Campana\altaffilmark{6,7}, Stefano Covino\altaffilmark{6} }

\altaffiltext{1}{Universit\'a degli studi di Roma ``Tor Vergata'', Via 
        della Ricerca Scientifica 1, I--00133 Roma, Italy; 
        nandad@mporzio.astro.it}
\altaffiltext{2}{INAF--Osservatorio Astronomico di Roma, Via Frascati 33, 
       I--00040 Monteporzio Catone (Roma),  Italy; 
       gianluca and stella@mporzio.astro.it}
\altaffiltext{3}{Astrophysics Missions Division, Research and Scientific 
         Support Department of ESA, ESTEC, Postbus 299, NL-2200 AG 
         Noordwijk, Netherlands; toosterb@rssd.esa.int }
\altaffiltext{4}{Istituto di Fisica Cosmica G. Occhialini, CNR,
Via Bassini 15, I--20133 Milano, Italy; sandro@mi.iasf.cnr.it}
\altaffiltext{5}{Laboratory of High Energy Astrophysics, Code 660.2,
        NASA/Goddard, Space Flight Center, MD 20771, USA;
        angelini@davide.gsfc.nasa.gov}
\altaffiltext{6}{INAF -- Osservatorio Astronomico di Brera, Via Bianchi
        46, I--23807 Merate (Lc), Italy; covino and campana@merate.mi.astro.it}
\altaffiltext{7}{Affiliated with ICRA}

\thispagestyle{empty}

\begin{abstract} 

We report the results of a long observation of the Anomalous X-ray
Pulsar \rxj\ obtained with the \BSAX\ satellite in August 2001.  The
best fit phase-averaged spectrum was an absorbed power law plus
blackbody model, with photon index $\Gamma \sim 2.4 $ and a black body
temperature of $kT_{bb} \sim 0.4 $\,keV. We confirm the presence of
significant spectral variations with the rotational phase of the
pulsar. In the spectrum corresponding to the rising part
of the pulse we found an absorption-like feature at $\sim
8.1$\,keV (a significance of 4$\sigma$), most likely due to cyclotron
resonant scattering. The centroid energy
converts to a magnetic field of $9\times10^{11}$\,G and
$1.6\times10^{15}$\,G in the case of electrons and protons,
respectively. If confirmed, this would be the first detection of a
cyclotron feature in the spectrum of an anomalous X-ray pulsar.

\end{abstract}


\section{INTRODUCTION}

Anomalous X-ray pulsars (AXPs) are characterized by spin periods in
the range of 5-12\,s, steady spin down ($\sim 10^{-11}\,ss^{-1}$),
steep and soft X--ray spectra with luminosities exceding by several
orders of magnitude their spin--down luminosities (Mereghetti
\& Stella 1995; van Paradijs, Taam \& van den Heuvel 1995).  All five
confirmed AXPs lie in the ``galactic'' plane and two (or three), 
are associated with supernova remnants. AXPs show no evidence
for a companion and are thus believed to be isolated neutron
stars either having extremely strong magnetic dipole fields ($\sim
10^{14}-10^{15}\,G$; ``magnetars''; Duncan \& Thompson 1992, Thompson
\& Duncan 1995) or accreting from a residual disk (Li 1999, 
Alpar 2001, Perna et al. 2000).  Alternatively, they can accrete from
a light companion. 1E\,1048--5937 and
1E\,2259+586, recently displayed short and intense
X--ray bursts (Graviil, Kaspi, \& Woods 2002; Kaspi \& Graviil
2002), strengthening a possible connection between AXPs and
Soft $\gamma$-Ray Repeaters (SGRs; Zhang 2002). 
For recent rewiews
see Israel, Mereghetti \& Stella 2002, Mereghetti et al. 2002, and
references therein.

\rxj\ was discovered with ROSAT (Voges et al. 1996); 
$\sim$\,11\,s pulsations were found in its X--ray flux
with ASCA (Sugizaki et al. 1997).  Early measurements 
suggested that is a fairly stable rotator (Israel et
al. 1999). In October 1999 a sudden spin--up event occurred,
which was interpreted as a glitch (Kaspi et al. 2000).  
Searches for an optical counterpart ruled out the presence of a
massive companion (Israel et al. 1999); an IR counterpart has
been recently proposed (Israel et al. 2003).  The diffuse ($\sim
8^{\prime}$) radio emission at 1.4 GHz (Gaensler et al. 2001) is due
to the supernova remnant G346.5--0.1, the association of which to the
AXP is still under debate.  There is no evidence of pulsed radio
emission from the AXP with an upper limit of 70$\mu$Jy on the pulsation
amplitude (Israel et al. 2002).  

Here we report the results of a long \BSAX\ observation of the AXP that
confirms significant spectral variations with pulse phase (Israel
et al. 2001) and shows the presence of an absorption-like feature
at $\sim$\,8.1 keV, probably due to cyclotron resonant scattering. 
 
\section{OBSERVATION}

The source was observed by \BSAX\ on 2001 August 17--22 with the
imaging Narrow Field Instruments (NFI): the Low Energy Concentrator
Spectrometer (LECS: 0.1--10\,keV; Parmar et al. 1997; 78.0\,ks of
effective exposure time) and the Medium Energy Concentrator
Spectrometer (MECS: 1.3--11\,keV; Boella et al. 1997; 200.0\,ks
exposure).  We extracted the source photons in the MECS and LECS
from a circular region of 6 arcmin radius around the position of the AXP. 
Photons extracted from a sourceless circular region of the same
size were used for background subtraction.  MECS and LECS photon
arrival times were corrected to the barycenter of the Solar System.

\section{RESULTS}

\subsection{{\it Timing and Spectral Analysis}}

In order to obtain a precise estimate of the pulse period, we divided
the MECS observation in nine time intervals and calculated the
pulsation phase for each of them. Fitting these phases with a linear
function gave a best period of $11.000563\pm 0.000005\,s$.
The folded light curve shows an energy-dependent profile
(Fig.\,1). Specifically, the pulse minimum shifts from a phase of
$\sim$\,0.0 in the lowest energy light curve (0.1--2\,keV) to
$\sim$\,0.3 in the 6--10\,keV light curve.  Correspondingly, the
pulsed fraction decreases from $\sim$\,30\,\% to $\sim$\,17\,\%.

We restricted the analysis of the LECS and MECS spectra to the
0.4--4\,keV and 1.65--10.8 \,keV energy range, respectively.  The 
spectra were binned so as to have about two bins per spectral
resolution element (FWHM). Futhermore, the data at the extremes of the
spectral range given above were further rebinned so as to have at
least 20 source events per bin (such that minimum chisquare techniques
could be reliably used in spectral fitting).

The spectra were well fit with an absorbed blackbody plus a power law
model (see Table\,1). The best fit of the phase-average spectrum 
gave a reduced $\chi^2$ of 0.95 for 298 degree of freedom (dof) 
for the following parameters: column
density of $ N_{H} = (1.36 \pm0.06)\times 10^{22}$\,cm$^{-2}$, a
blackbody temperature of $kT_{bb}= 0.44\pm0.01$\,keV (blackbody
radius of $R_{bb} = 6.6\pm0.4$\,km, assuming a distance of 5\,kpc)
and a photon index of $\Gamma = 2.40\pm0.06$ (all error bars in the text
are 90\,\% confidence). The unabsorbed flux 
in the 0.5--10\,keV range was $1.87\times
10^{-10}$\,erg\,cm$^{-2}$\,s$^{-1}$ corresponding to a luminosity of
5.6$\times10^{35}$\,erg\,s$^{-1}$ (for a 5 kpc distance). In
the 0.5--10 keV band the blackbody component accounts for
$\sim$\,30\,\% of the total unabsorbed flux. We also tried other
combinations of spectral models (cutoff power law plus blackbody or 
two blackbodies with different temperatures) but they all produced larger
reduced $\chi^2$ values.

\subsection{{\it Pulse Phase Spectroscopy}}

We adopt for zero phase the minimum of the pulse profile in the
0.2--3\,keV folded light curve (see Fig.\,1). In order to carry out
pulse--phase spectroscopy we accumulated spectra in six
different phase intervals. The boundaries of these (0.0, 0.26, 0.4,
0.58, 0.7, 0.84, 1.0) were designed so as
to sample separately the minimum, rising,
maximum, and decaying part of the pulse
profile, while maintaining a sufficiently good statistics for a
detailed spectral study (results in Tab.\,1, Fig.\,2, 3 and 4). 
A significant variation of the spectral parameters with pulse
phase was clearly seen (especially for $\Gamma$ and $R_{bb}$, see Fig.\,2; 
see also Israel et al. 2001). In all intervals but one, an acceptable fit was obtained
with the absorbed power--law plus black body model (reduced $\chi^2$
in the 0.8--1.1 range); a reduced $\chi^2$ of
1.2 was instead obtained in the 0.4--0.58 phase
interval. Specifically, the data were systematically below the best
fit model in the $\sim$\,7.8--8.4\,keV range (see Fig.\,4b). We tried to fit
three different models: a Gaussian, an absorption edge and a cyclotron
feature.  While the inclusion of a Gaussian or an absorption edge did
not lead to a significant improvement of the fit, the cyclotron model
(CYCLABS in the XSPEC package, see Mihara et al. 1990 for details) 
led to a reduced chisquare of 1.1 for
86 dof, corresponding to an F-test probability of $3\times10^{-3}$
(3.1$\sigma$).

In order to improve the statistics we added in phase the spectra from
the 1999 \BSAX\ observation of the source (Israel et
al. 2001), increasing the total exposure time to $\sim$ 250\,ks.  For
this we used the pulse period determination from the timing solution
of a 4 year--long \RXTE\ monitoring (Kaspi et
al. 2000)
\footnote {Note that we could not use for the 2001 dataset the 
post--``glitch'' timing (Gavriil \& Kaspi 2002) 
since the period extrapolated from their parameters significantly differs 
from that derived here ($\Delta P \simeq 1.5\times 10^{-3}$ s)}
. The determination of the zero phase was done again by using the 
pulse profile minimum in the 0.3--2\,keV energy range (note that no
significant shape difference was found with respect to the 2001
observation).  We estimate that our procedure can introduce an
uncertainty of up to 0.01 in the phasing of the two
observations which is negligible for the aims of our study.

For each of the six phase intervals, a spectrum was summed 
together with the corrisponding spectrum from the 2001
observation. The spectral analysis was then repeated.
In the MECS and LECS spectra from the 0.4--0.58
phase interval the inclusion of a Gaussian ($E_{g}$\,=\,8.3\,keV, 
$\sigma$\,=\,0.4\,keV 
and $\chi^2$\,=\,1.1) and an absorption edge ($E_{e}$\,=\,7.7\,keV, 
$\tau$\,=\,0.4 and $\chi^2$\,=\,0.9)  to fit
the feature at $\sim 8.1$\,keV resulted in an improvement of the
chisquare, which converted to a single-trial F-test probability 
of 0.18 ($\sim 1\sigma$) and $4\times10^{-4}\,(\sim 3.7
\sigma$), respectively. A much more significant improvement 
was obtained by adding instead a Resonant Cyclotron Feature (RCF)
model; the F-test probability in
this case was $1.8\times10^{-5}$ corresponding to a single trial
significance of 4.5$\sigma$ or 4$\sigma$ after correction for the six
spectra that we analysed (see Fig.\,4 and Tab\,1).

\section{DISCUSSION}

During a \BSAX\ study of \rxj\, we
discovered an absorption-like feature at an energy of $\sim$\,8.1\,keV
in a pulse phase interval corresponding to the rising part of the
$\sim$\,11\,s pulse. This feature was best fit by a resonant cyclotron
feature model with a centroid energy of $\sim$8.1\,keV and an 
equivalent width of $\sim$460\,eV.

The detection of an RCF in a specific pulse-phase interval
and superposed to an X-ray
continuum that varies with the pulse phase is reminiscent of the
behaviour seen in standard accreting pulsars in X-ray binaries (see
Wheaton et al. 1979, Santangelo et al. 1999). 
If interpreted as an electron
resonant feature at the base of the accretion column, the feature at
$\sim$\,8.1 keV implies a neutron star surface magnetic field of
$\sim 9.2 \times 10^{11}$\,Gauss (using a gravitational
redshift z=0.3) . This value is just slightly lower
than that measured for electron RCFs in typical accreting X-ray
pulsars (see Fig.\,5); more interestingly it is close to that
required by models for AXPs which involve residual disk accretion in the 
spin--down regime (Mereghetti \& Stella 1995; Alpar 2001; Perna
et al. 2001 ). In this context, one can solve the torque equation
(see e.g. Eq. 11.35 in Henrichs 1983) by exploiting the measured value
of $\dot{P}$ and range of accretion luminosity derived from
plausible distances (5--10\,kpc). The surface magnetic (dipole) field
obtained in this way is 0.6--1.1$\times10^{12}$\,G (corrisponding to a
fastness parameter range of $\omega_s = 0.57-0.54$, a typical value
for the spin-down accretion regime; see Ghosh \& Lamb 1979 and Henrics 1983).

The agreement of this estimate with the magnetic field inferred from
the electron RCF interpretation is intriguing, especially in consideration 
of the other analogies
with the pulse-phase spectral dependence of conventional accreting
X-ray pulsars. By contrast, if an electron RCF
arose somehow at the polar caps of a rotation powered pulsar, a
B--field strength of 9.2$\times10^{11}$\,G would be in the range 
of many radio pulsars and yet much lower than that required
to spin--down at the observed rate through magnetic dipole radiation
($\sim 5 \times 10^{14}$\,G, indeed this was one of the motivations for
magnetar model, see below).

Alternatively the RCF might be due to protons.
For the magnetic field strengths forseen in the ``magnetar''
scenario, proton cyclotron features (if any) are expected to lie in
the classical X--ray band (0.1--10\,keV; Zane et al. 2001; Lai \& Ho 2002).  
A proton RCF feature at $\sim 8.1$\,keV would correspond to surface field of
1.6$\times$10$^{15}$\,G (z=0.3). The fact that this
value is $\sim 3$ times  higher than the surface field derived
from the usual magnetic dipole spin down formula should not be of
concern. According to the magnetar model, the magnetic field
at the star surface and its vicinity is dominated by higher order
multipole field components
(Thompson \& Duncan 1995). At large radii
the dipole component, responsible for the secular spin--down,
dominates.  It is thus expected that a proton RCF feature, sampling
the (total) surface magnetic field strength, provides a higher value
than the mere dipole component.

There is a clear correlation between the width and centroid energy of
the electron RCFs in accreting X--ray pulsars (see Fig.\,5 extending
the results of Orlandini \& Dal Fiume 2001). The values from \rxj\ and
SGR\,1806--20 are in good agreement with such a relation. The modest
range of width to centroid energy ratio implied by this indicates that
magnetic field geometry effects at the neutron star surface likely
dominate the RCF width (on the contrary temperature and particle mass
would alter this ratio).  This, in turn, suggests that similar
(relative) ranges of surface magnetic field strength are ``sampled''
by RCFs in accreting X--ray pulsar and, by extension, RCFs in AXPs and
SGRs.

Other interpretations of the feature at $\sim 8.1$~keV appear less
likely. Firstly, fitting an edge or line due to photo-electric absorption
provides a less pronounced improvement of the fit than the RCF model. 
Secondly, an edge by iron
at a sufficiently large distance from the neutron star that energy
shifts are negligible would require a high overabundance of this
element and intermediate ionisation stages (such as C--like iron). Yet
it has long been known that the photoionisation equilibrium of such a
plasma is unstable (Krolik \& Kallman 1984; Nagase 1989). 
The energy of an ion feature forming
in the neutron star atmosphere would be drastically altered by
magnetic field effects (see Mori \& Hailey 2002 and references
therein). In this and the above interpretations, however, it would
also be difficult to explain why an ion feature is observed only over
a restricted range of pulse phases. Exploring in detail these
possibilities is beyond the scope of this letter.

In conclusion, we found an absorption-like feature in the \BSAX\
X--ray spectrum of \rxj\ taken during the rising phase of the
$\sim$11\,s pulse, which is best fit by a cyclotron resonant
scattering model. If this interpretation is correct, the centroid
energy translates into a magnetic field strength of
$\sim$1.6$\times$10$^{15}$\,G and $\sim$9.2$\times$10$^{11}$\,G
depending on whether protons or electrons, respectively, are
responsible for the feature.

\acknowledgments
Nanda Rea acknowledges useful discussions with  S. Dall'Osso.  
This work was partially supported through ASI and COFIN 2000 grants.

\clearpage


\input{tab1.tex}


\clearpage

\begin{figure}
\plotone{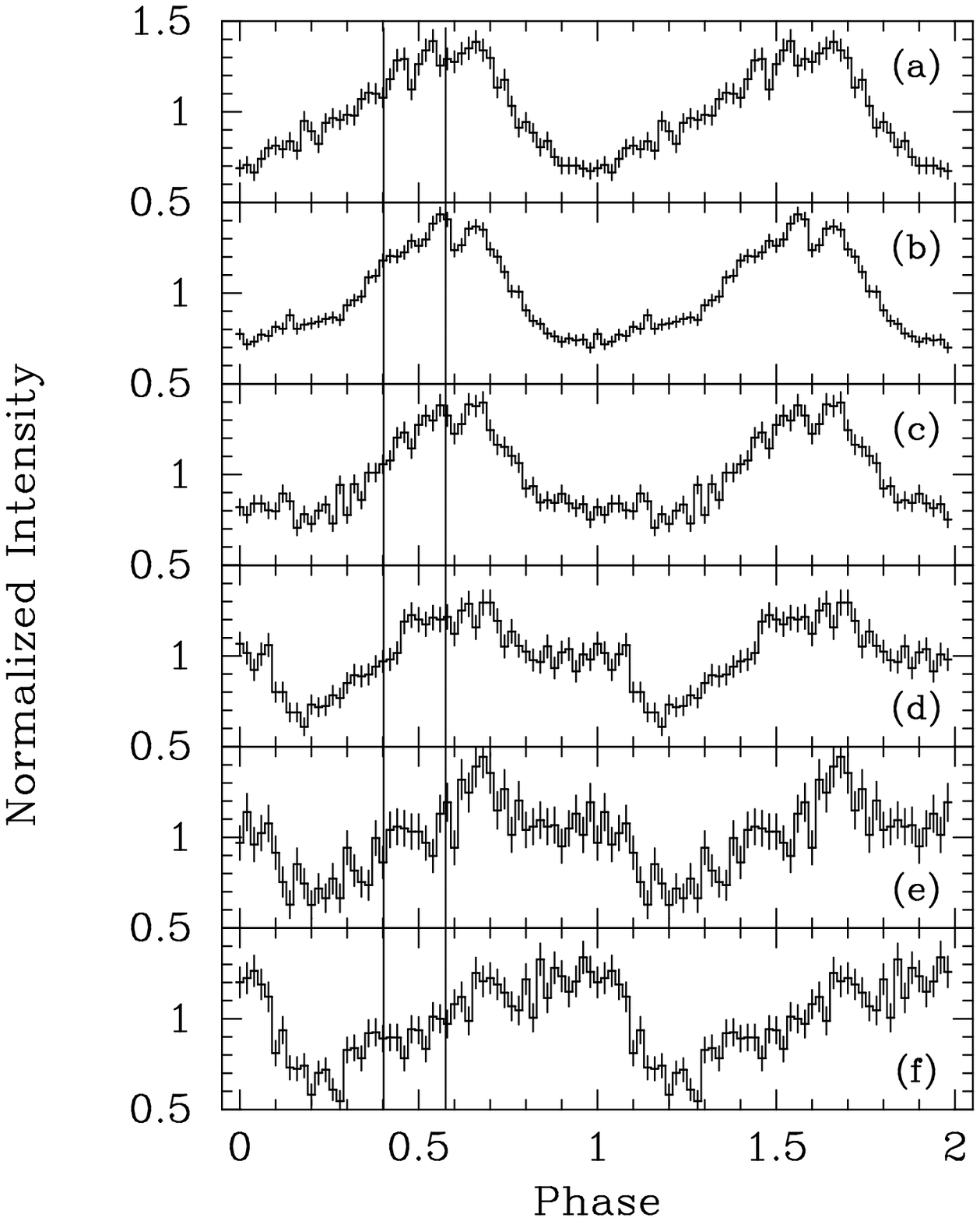}
\caption{MECS light curves of \rxj\ folded at the best spin period 
(two pulse cycles are shown) 
for six energy bands: (a) 0.1--2 keV; 
(b) 2--3 keV; (c) 3--4 keV; 
(d) 4--5 keV; (e) 5--6 keV; (f) 6--10 keV. The vertical lines mark the 
phase interval in which the absorption-like feature was detected. 
The data in Figures 1-3 are from the 2001 \BSAX\  observation \label{fig1}}
\end{figure}


\begin{figure}
\plotone{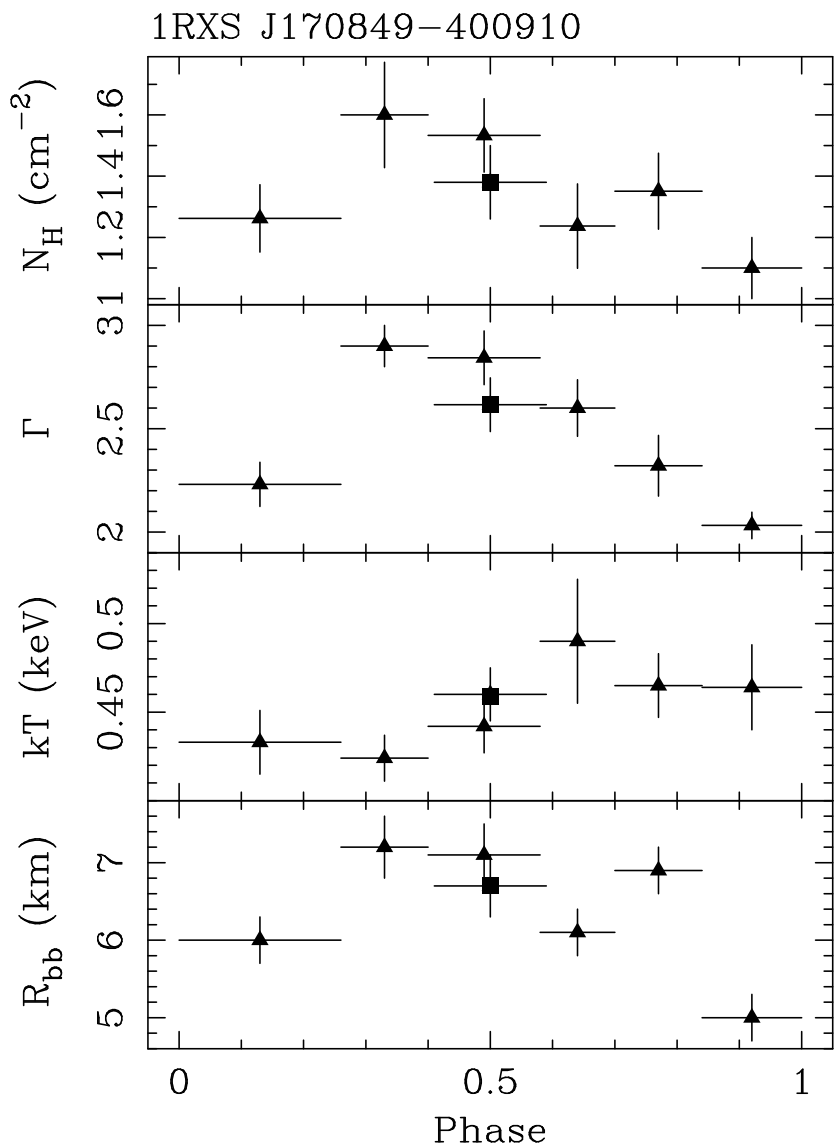}
\caption{Spectral parameter variability from pulse-phase spectroscopy using 
 an absorbed
  black body plus power-law model.  Filled squares represent the spectral
  parameters after the addition a cyclotron line in the 0.4 -- 0.58 phase 
  interval. Note that the relevant points are slightly
  shifted in phase for clarity.  The error bars are 
  $1\sigma$ . \label{fig2}}
\end{figure}


\begin{figure}
\plotone{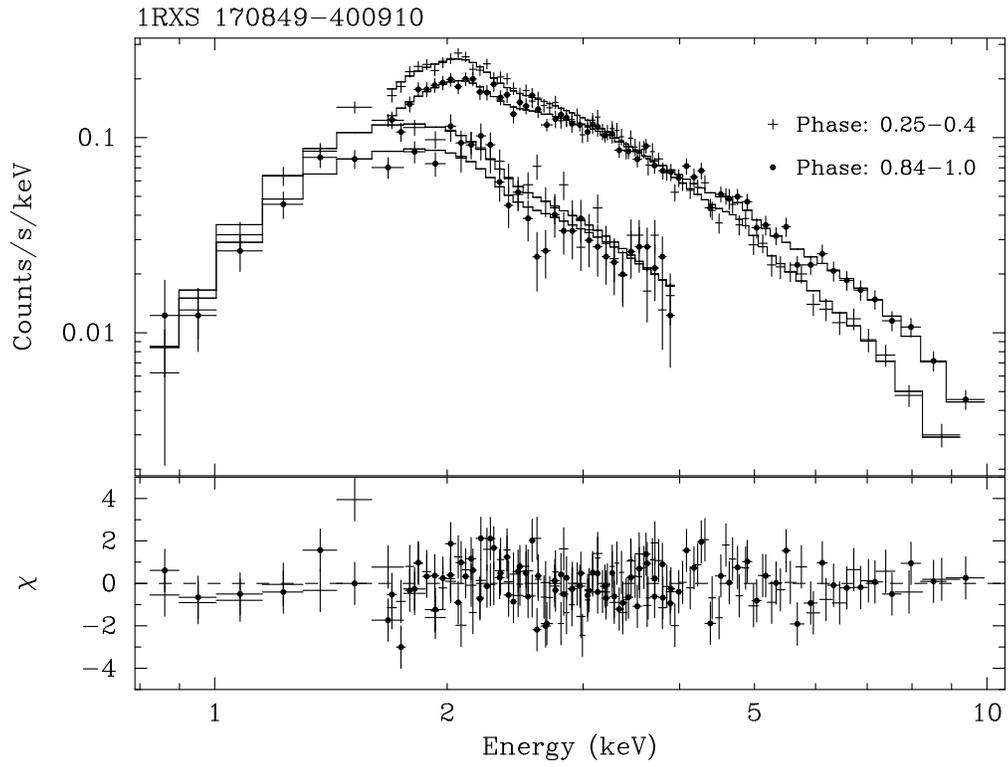}
\caption{LECS and MECS spectra in the two phase intervals 
 characterised by the highest and lowest value of the power 
 law photon index $\Gamma$; see also Table 1.\label{fig3}}
\end{figure}


\begin{figure}
\plotone{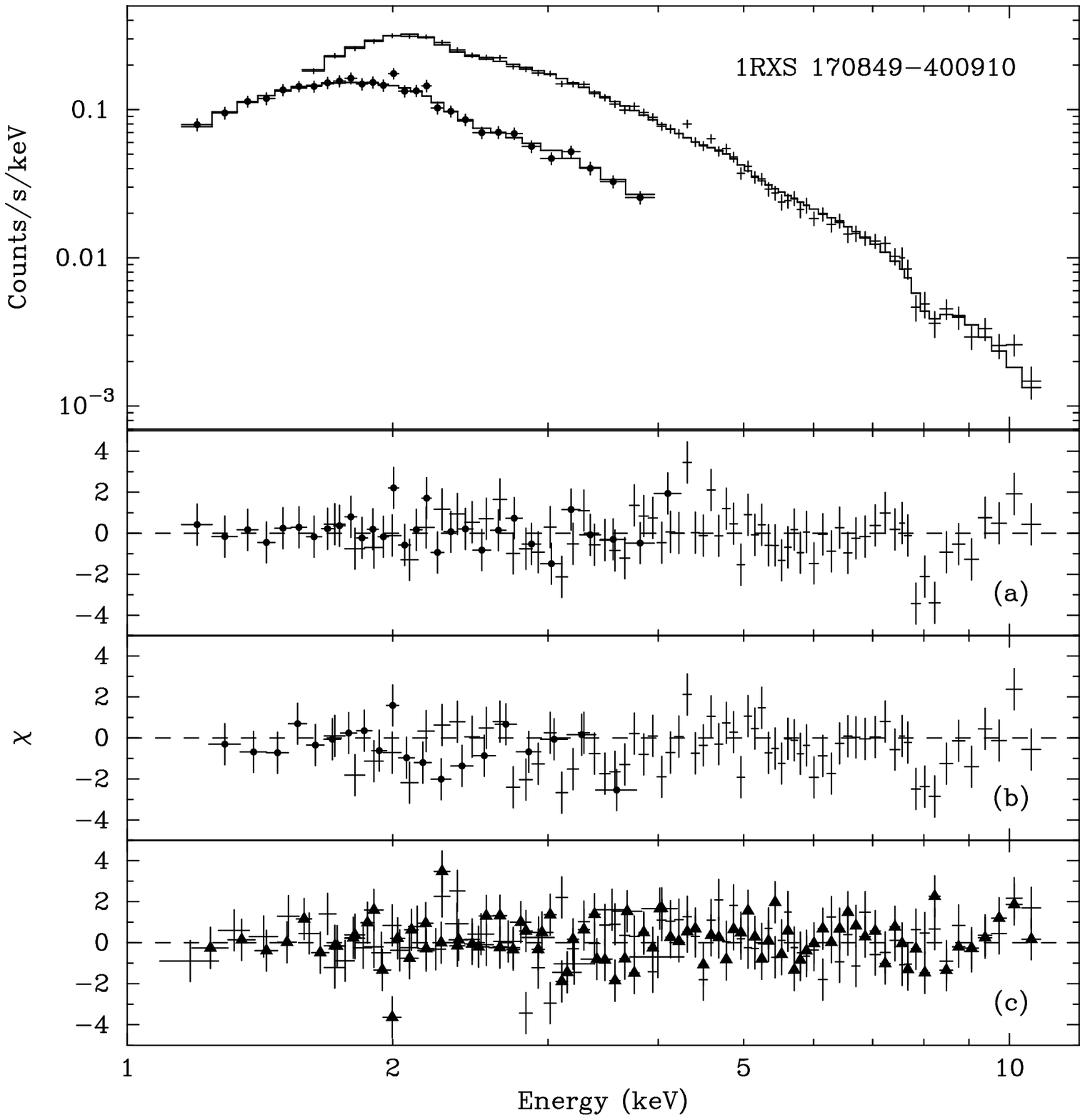}
\caption{MECS and LECS spectra from the 0.4--0.58 
 phase interval fit with the ``standard model'' 
(the sum of a blackbody and power law with absorption) plus a cyclotron line.
 Residuals are relative to the ``standard model' alone in order to
 emphasize the absorption-like feature at $\sim 8.1$~keV: 
 (a) the \BSAX\ observations
 merged together; (b) the 2001 observation alone; (c) the phase
 intervals contiguous to that showing the cyclotron absorption feature 
 in the merged observations. \label{fig4}}
\end{figure}


\begin{figure}
\plotone{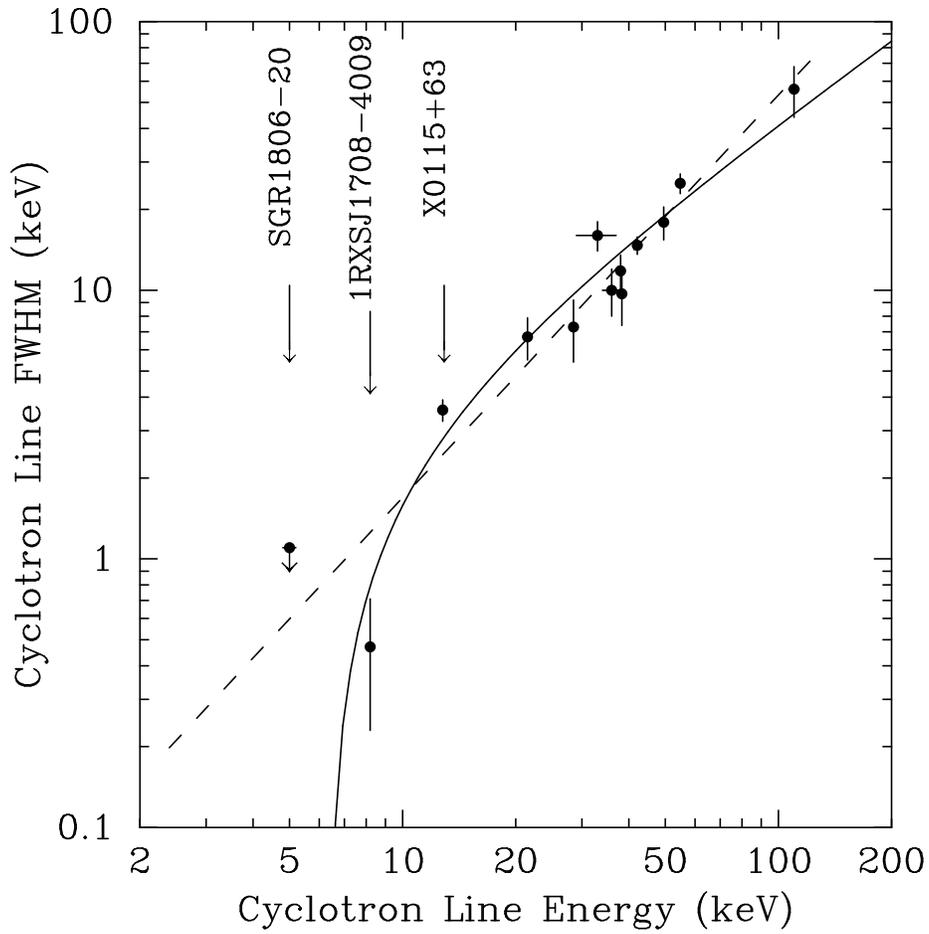}
\caption{Line width vs. centroid energy from a sample of 
          accreting X-ray pulsars
          with electron RCFs (Orlandini \& Dal Fiume 2001), \rxj\,
         (this Letter) and SGR\,1806--20 (Ibrahim et al. 2002). The
         dashed and solid lines give the best fit models based on a
         power law model (slope of $\sim 1.5$) and a constant plus a
         linear function, respectively.}
\label{fig5}
\end{figure}


\end{document}

%% file: tab1.tex


\begin{table}
\begin{center}
\caption{
  Best fit spectral parameters from selected pulse-phase intervals:
that showing feature at $\sim$8.1\,keV (0.4-0.58) and those
characterized by the lowest and the highest power law photon index
$\Gamma$ (see also Fig.\,3).
All fluxes are unabsorbed and calculated in 0.5--10\,keV range;
uncertainties are 90\% confidence.} \vspace{0.3cm}
\begin{tabular}{lccc}
\hline 
\hline
&  &  Phase Intervals & \\ 
Parameters    & 0.4--0.58 & 0.84--1.0 & 0.26--0.4 \\ 
\hline
& & & \\

$ N_{H}(\times10^{22}$\,cm$^{-2})$ & $1.38\pm0.01$ 
                        & $1.1\pm0.1$ & $1.6\pm0.3$ \\ 
$\Gamma$                         & $2.6\pm0.1$   
                        & $2.0\pm0.1$ & $2.9\pm0.1$ \\
PL flux ($\times 10^{-11}$erg\,cm$^{-2}$\,s$^{-1}$)    & 21. 
                        & 8.3 & 24.8  \\
 & & & \\
$kT_{bb}$ (\,keV)            & $0.46\pm0.01$     
                        & $0.46\pm0.02$ & $0.42\pm0.01$ \\                   
 BB flux  ($\times 10^{-11}$\,erg\,cm$^{-2}$\,s$^{-1}$) & 9.1   
                        & 5.1 & 6.1 \\
$R_{bb}$ (d = 5\,kpc; km)  & $6.7\pm0.4$      
                        & $5.0\pm0.6$ & $7.2\pm0.6$ \\
& & & \\
$E_{cyclabs}$ (\,keV)        & $8.1 \pm 0.1 $ & ... & ... \\
Line Width  (\,keV)          & $0.2\pm 0.1$     & ... & ... \\
Line Depth  (\,keV)          & $0.8 \pm 0.4$    & ... & ... \\
& & & \\
$ \chi^2/d.o.f.  $                    &  1.09 
                         & 1.01  & 0.98  \\
Flux ($\times 10^{-11}$\,erg\,cm$^{-2}$\,s$^{-1}$)     & 30.0 
                         & 7.3   & 7.2  \\
\hline
\hline
\end{tabular}
\end{center}

\end{table}


%% file: ms.bbl
\clearpage

\begin{thebibliography}{}


\bibitem {1}
       Alpar, M.A., 2001, ApJ, 554, 12-45
\bibitem {2}
       Boella, G., \etal\ 1997, A\&AS, 122, 327
\bibitem {3} 
       Duncan, R.C., \& Thompson, C. 1992, ApJ, 392, L9
\bibitem {4}
         Gaensler, B.M., et al. 2000, MNRAS, Volume 318, Issue 1, pp.58-66     
\bibitem {5}
         Ghosh, P., Lamb, F. K., 1979, ApJ, Part 1, vol. 234, p. 296-316
\bibitem {6}
         Graviil, F.P. \& Kaspi, V.M. 2002, ApJ, 567, 1067G 
\bibitem {7}  
         Graviil, F.P., Kaspi V.M., \& Woods, P.M. 2002, Nature, 419, 142G
\bibitem{8}
         Henrichs, H.F., 1983 Accretion-driven stellar X-ray sources P.419
\bibitem {9} 
         Ibrahim, A.I., Swank, J.H., Parke, W. 2002, astro--ph, 0210515
\bibitem {10} 
        Ibrahim, A.I., Safi-Harb, S., Swank, J.H., Parke, W.,
         Zane, S., Turolla, R. 2002, ApJ, 574, L51 
\bibitem {11}
        Israel, G.L., Covino, S., Stella, L., Campana, S., Haberl, F.,
         Mereghetti, S. 1999, ApJ, 518, L107
\bibitem {12}
        Israel, G.L., Oosterbroek, T., Stella, L., Campana, S., 
         Mereghetti, S., Parmar, A., 2001, ApJ, 560, L65
\bibitem {13} 
         Israel, G.L., Mereghetti, S., \& Stella, L. 2002a ,
        in  $\gamma$--Ray Bursts in the Afterglow Era, ed. S. Mereghetti \&
        M. Feroci, Mem.S.A.It., Vol.\,73, N.\,2, pag.\,465   
\bibitem {14}
        Israel, G.L., et al. 2003, ApJ submitted
\bibitem {15}
        Kaspi, V.M., Lackey, J.R., Chakrabarty, D. 2000, ApJ, 537, L31
\bibitem {16}
        Kaspi, V.M., \& Gravill, F.P., 2002, IAU Circ, No.7924
\bibitem {17} 
        Krolik, J.H., \& Kallman, T.R. 1984, ApJ, 286, 366K 
\bibitem {18}
        Lai, D., \& Ho, W.C.G., 2002, astro--ph 0211315  
\bibitem {19}
        Li, X.-D., 1999, ApJ, 520, 271L
\bibitem {20}
        Mereghetti, S., \& Stella, L. 1995, ApJ, 442, L17 
\bibitem {21}
        Mereghetti, S., Chiarlone, L., Israel, G.L., Stella, L. 2002, MPE Rep, 278; 
         Garching: MPE, 29 
\bibitem {22} 
        Mihara, T., et al., 1990, Nature, 346, 250M  
\bibitem {23}
        Mori, K., \& Hailey, C. J. 2002, ApJ, 564, 914 
\bibitem {24}
        Nagase, F. 1989, PASJ, 41, 1-79
\bibitem {25}
        Orlandini \& Dal Fiume  2001, AIP conference proceedings 1999, Vol.599 p.283 
\bibitem {26}
        Parmar, A.N., et al. 1997, A\&AS 122, 309 
\bibitem {27}
        Perna, R., Heyl, J., Hernquist, L., Juette, A., Chakrabarty, D. 2001, ApJ, 557, 18P
\bibitem {28}
        Perna, R., et al., 2000, ApJ, 541, 344 
\bibitem {29}
        Santangelo, A., et al., 1999, ApJ, 523, L85 
\bibitem {30}
        Sugizaki, M., et al , 1997, PASJ, v.49, p.L25-L30
\bibitem {31}
        Thompson, C., \& Duncan, R.C. 1995, MNRAS, 275, 255
\bibitem {32}
        van Paradijs, J., Taam, R.E., \& van den Heuvel, E.P.J., 1995, 
           A\&A, 299, L41 
\bibitem {33}
        Voges, W., et al. 1996, IAU Circ., 6420, 2 (1996). 
        Edited by Green, D. W. E. 
\bibitem {34}
        Wheaton, W.A., et al., 1979, Nature 282, p.240
\bibitem {35}
        Zane, S., Turolla, R., Stella, L., \& Treves, A., 2001, \apj, 560, 384
\bibitem {36}
        Zhang, B., 2002, astro-ph, 0212016

\end{thebibliography}
